\documentstyle[12pt]{article}
\begin{document}

\title{Theory of hybrid systems. II. The symmetrized product and
redefined Lie bracket of quantum mechanics}
\author{Slobodan Prvanovi\'c and Zvonko Mari\'c \\
{\it Institute of Physics, P.O. Box 57, 11080 Belgrade,} \\
{\it Yugoslavia }}
\date{}

\maketitle
\begin{abstract}
The symmetrized product for quantum mechanical observables is
defined. It is seen as consisting of the ordinary multiplication
and the application of the superoperator that orders the operators
of coordinate and momentum. This superoperator is defined in a way
that allows obstruction free quantization when the observables are
considered from the point of view of the algebra. Then, the
operatorial version of the Poisson bracket is defined. It is shown
that it has all properties of the Lie bracket and that it can
substitute the commutator in the von Neumann equation. 
\end{abstract}

\section{Introduction}

The Hermitian operators $\hat q$ and $\hat p$, acting in the
Hilbert space of states, represent coordinate and momentum of
quantum mechanical system and correspond to the coordinate $q$ and
momentum $p$ of classical mechanical system. For these operators
it should hold $[\hat q , \hat p ] =i\hbar \hat I$. This is one
part of the quantization procedure other parts of which are the
following. One has to say which operator should correspond to, let
say, $q^n p^m$ of classical mechanics. That is where the problem
of symmetrized product (ordering rule) of quantum mechanics
firstly enters. Namely, one is not willing to accept $\hat q ^n
\cdot \hat p ^m $ because this operator, in contrast to the
involved ones, is not the Hermitian. (With $\cdot$, which will be
sometimes omitted, we represent ordinary product - successive
application, of operators.) One has to define new operation,
denoted by $\circ$, according to which the product of $\hat q ^n$
and $\hat p ^m$ will be the Hermitian. The most convenient way for
doing that is by using already existing operations of addition,
ordinary multiplication  and multiplication by numbers. With this
new operation the algebraic structure of quantum mechanics appears
in its rudimentary form. In Cohen (1966) and Kerner and Sutcliffe
(1970) one can find a short review of different propositions for
symmetrized product while in Shewell (1959) there is a critical
discussion of many ordering rules introduced in there cited
references.

On the other hand, the classical mechanical variables have structure of
both the algebra and Lie algebra. The Lie product, or bracket, of
classical mechanics is the Poisson bracket:
\begin{equation}
\{ f(q , p ), g(q , p ) \} = {\partial f( q , p )\over \partial q
}\cdot {\partial g(q , p )\over \partial p } - {\partial g(q , p )\over
\partial  q }\cdot {\partial f( q , p )\over \partial p },
\end{equation}
where $f( q , p )$ and $g( q , p )$ belong to the set of classical
variables. Its importance comes from the fact that it appears in the
dynamical equation (as well as in all transformations due to
symmetries). The expression:
\begin{equation}
{\partial \rho (q,p,t) \over \partial t }= \{ H(q,p), \rho (q,p,t) \},
\end{equation}
is the Liouville equation where $\rho (q,p,t)$ is distribution
representing state and $H(q,p)$ is the Hamiltonian of classical system.

It is desired to equip the set of quantum mechanical observables not
just with the algebraic (symmetrized) product, but with the Lie
algebraic product as well, {\it i.e.}, one has to introduce the quantum
analogue of the Poisson bracket. After Dirac (1958), the canonical
quantization prescription says that the Poisson bracket of classical
variables has to be translated into the ${1\over i\hbar }$ times the
commutator of corresponding quantum observables. The most important
place where the commutator appears is the von Neumann equation:
\begin{equation}
{\partial \hat \rho (t) \over \partial t }=
{1\over  i\hbar }[ H(\hat q, \hat p), \hat \rho (t) ].
\end{equation}
It prescribes the way in which the state of quantum system $\hat
\rho (t)$ evolves due to the Hamiltonian $H(\hat q, \hat p)$.
(Similar expressions hold for all symmetry transformations.) The
last equation, of course, for a pure state ($\hat \rho (t) ^2 =
\hat \rho (t)$) reduces to the Schr\"odinger equation.

Discussions on the algebraic and Lie algebraic structures in
quantum mechanics one can find in Emch (1972) and Joseph (1970). In
Arens and Babit (1965), Gotay (1980) and (1996) and Chernoff (1995), it
was found that these two are interrelated in such a way that there is
an obstruction to quantization which is manifested through the
existence of some contradiction. Concretely, if the Lie  bracket of
quantum observables $f_1 (\hat q , \hat p )$ and $f_2 (\hat q , \hat p
)$ is not equal to the Lie bracket of $f_3 (\hat q , \hat p )$ and
$f_4 (\hat q , \hat p )$, while for the Poisson brackets of
classical variables, to which these quantum observables should
correspond, \ holds \ $\{ f_1 (q,p) , f_2 (q,p) \} \ = \ \{ f_3 (q,p) ,
f_4 (q,p) \}$, then there are two different operators that can be
attributed to the same function and quantization can not  be taken
as selfconsistent. One can conclude, see Chernoff (1995), that the
problem of quantization, which we have sketched above, is impossible.

This article is the second in series devoted to the development of
theory of hybrid systems. Within it we shall define the
symmetrized product of quantum mechanics at the first place. Then,
by using this product, we shall be able to propose new Lie bracket
of quantum mechanics. It will be the operatorial version of the
Poisson bracket. In this way the algebraic and Lie algebraic
aspect of quantum and classical mechanics will appear to be the
same, as will become obvious latter. This result is of great
importance for the foundation of theory of hybrid systems.
However, since the ordering rule and Lie bracket are {\it per se}
interesting and deserve detailed consideration, we shall approach
them here without reference to the rest of theory of hybrid
systems.

\section{Definition of the symmetrized product}

The most often used example of symmetrized product of two
operators is ${1\over 2} (\hat q \hat p + \hat p \hat q )$. This
operator is the Hermitian and it corresponds to $qp$ of classical
mechanics. But, this  expression is susceptible of more than one
meaning. Namely, one can look on ${1\over 2} (\hat q \hat p + \hat
p \hat q )$ as on ``one half of the anti-commutator of $\hat q$
and $\hat p$" or one may take it as ``the sum of all combinations
of involved operators divided by the number of this combinations".
Different understandings of this generally accepted expression
lead to constructions of nonequivalent expressions in other, not
so trivial cases. In Kerner and Sutcliffe (1970) it was shown that
there given propositions for symmetrized product start to differ
for quartic monomials in $\hat q$ and $\hat p$.

We believe that by examining the way in which ${1\over 2}(\hat q
\hat p + \hat p \hat q )$ was constructed from $\hat q$ and $\hat
p$, that are basic elements of the algebra, one gets better
insight in how to solve the ordering problem in general. The
procedure could be: $\hat q$ and $\hat p$ were multiplied as they
were ordinary numbers, resulting in $\hat q \hat p$, and then some
ordering procedure was applied, producing ${1\over 2}(\hat q \hat
p + \hat p \hat q )$. Taken in this way, the symmetrized product
becomes a two step operation. The first step of symmetrized
product is multiplication as in the case of c-numbers, with all
well known properties except commutativity. The second step is the
ordering procedure. We see it as the application of some
superoperator ${\bf S}$ on the operatorial expressions consisting
of $\hat q$'s and $\hat p$'s. Then, definition of the ordering
procedure becomes a definition of superoperator ${\bf S}$ that
acts on sequences of $\hat q$ and $\hat p$ assigning them, in some
well defined manner, another sequences. By adopting this way of
looking on symmetrized product, as we are going to do in what
follows, it may be said that propositions of ordering rules given
in literature differ since they (implicitly) use different
superoperators - symmetrizers.

We shall use ${\bf S}$ defined as the linear superoperator that
acts on sequences of $\hat q$ and $\hat p$ producing the sum of
all different combinations of that operators and dividing the
result by the number of these combinations. For example, ${\bf
S}(\hat q ^2 \hat p ^2
 )= {1\over 6}(\hat q ^2 \hat p ^2 + \hat q \hat p \hat q \hat p + \hat
q \hat p ^2 \hat q + \hat p \hat q ^2 \hat p + \hat p \hat q \hat p
\hat q + \hat p ^2 \hat q ^2 )$. In general, when ${\bf S}$ acts on
sequence $\hat q \hat p \hat p \cdots \hat p \hat q \hat q$, where the
operator of coordinate appears $n$ times and the operator of momentum
appears $m$ times, no matter how they are ordered in this sequence, the
result will be:
\begin{equation}
{n!m! \over (n+m)!} (\hat q ^n \hat p ^m + \cdots + \hat p ^m \hat q ^n
 ).
\end{equation}
It is understood that in the parenthesis there should be all
different combinations of $n+m$ operators. Since there are $n$
operators of the one kind and $m$ of the other, there should be
${(n+m)!\over n!m!}$ terms in the parenthesis.

More formally, the action of ${\bf S}$ on operatorial sequences is
defined by:
\begin{equation}
{\bf S}(\hat q ^{a_1} \hat p ^{b_1} \cdots \hat q ^{a_n} \hat p
^{b_n})={(\sum _i a_i )! (\sum _i b_i )! \over (\sum _i a_i +\sum _i
b_i )!} \sum ' _{{{c_1 , \cdots , c_m \atop d_1 , \cdots , d_m } \atop
\sum _j c_j = \sum _i a_i } \atop \sum _j d_j = \sum _i b_i} \hat q ^{c_1}
\hat p ^{d_1} \cdots \hat q ^{c_m} \hat p ^{d_m},
\end{equation}
where $a_i , b_i , c_j , d_j \in {\bf N }_o , \ i\in \{ 1, \cdots
, n \}, j\in \{1,\cdots , m \}, \ n\in {\bf N} ,\ m=\sum _i a_i +
\sum _i b_i +1$ and where the prime over the sum indicates the absence
of repeated combinations, {\it i.e.}, if:
$$
\hat q ^{c'_1} \hat p ^{d'_1} \cdots \hat q ^{c'_m} \hat p ^{d'_m}=
\hat q ^{c''_1} \hat p ^{d''_1} \cdots \hat q ^{c''_m} \hat p ^{d''_m},
$$
then $c'_j =c''_j$ and $d'_j =d''_j$ for all $j$. (The immediate
consequence of (5) is that:
\begin{equation}
{\bf S}(\hat q ^{a_1} \hat p ^{b_1} \cdots \hat q ^{a_n} \hat p
^{b_n})={\bf S}(\hat q ^{c_1} \hat p ^{d_1} \cdots \hat q ^{c_m} \hat p
^{d_m}),
\end{equation}
whenever $\sum _i a_i = \sum _j c_j$ and $\sum _k b_k = \sum _l d_l$.)
The linearity of $\bf S$ reads:
$$
{\bf S}(k \hat q ^{a_1} \hat p ^{b_1} \cdots \hat q ^{a_n} \hat p
^{b_n} +  l\hat q ^{c_1} \hat p ^{d_1} \cdots \hat q ^{c_m} \hat p
^{d_m})=
$$
\begin{equation}
=k{\bf S}(\hat q ^{a_1} \hat p ^{b_1} \cdots \hat q ^{a_n} \hat
p ^{b_n})+ l{\bf S}(\hat q ^{c_1} \hat p ^{d_1} \cdots \hat q ^{c_m}
\hat p ^{d_m}),
\end{equation}
where $a_i , b_i , c_j , d_j \in {\bf N}_o , \ i\in \{ 1, \cdots , n
\}, j\in \{1,\cdots , m \}, \  n, m\in {\bf N} , \  k, l \in {\bf C}$.
The third defining property of $\bf S$ is:
\begin{equation}
{\bf S}(k \hbar ^m \hat q ^{a_1} \hat p ^{b_1} \cdots \hat q ^{a_n}
\hat p ^{b_n})=0,
\end{equation}
where $a_i , b_i \in {\bf N}_o , \ i\in \{ 1, \cdots , n \}, \  n\in
{\bf  N} , k\in {\bf C} , \  m\in {\bf N} $. Necessity of this property
will be discussed in the next section.

For reasons that will be given in Sec. 4, the symmetrized product of
observables and the partial derivatives of state $\hat \rho$ is defined
in way which is a slight modification of the above. Precisely, (5)
becomes:
$$
{\bf S}(\hat q ^{a_1} \hat p ^{b_1} ({\partial \hat \rho \over
\partial \hat r })^{e_1} \cdots \hat q ^{a_n} \hat p ^{b_n} ({\partial
\hat \rho \over \partial \hat r })^{e_n})=
$$
\begin{equation}
={(\sum _i a_i )! (\sum _i b_i )! \over (\sum _i a_i +\sum _i b_i +
\sum _i e_i)!} \sum  ' _{{{{{c_1 , \cdots , c_m \atop d_1 , \cdots ,
d_m } \atop f_1 , \cdots , f_m }  \atop \sum _j c_j = \sum _i a_i }
\atop \sum _j d_j = \sum _i b_i } \atop \sum _j f_j =  \sum _i e_i }
\hat q ^{c_1} \hat p ^{d_1} ({\partial \hat \rho \over \partial \hat r
})^{f_1} \cdots \hat q ^{c_m} \hat p ^{d_m} ({\partial \hat \rho \over
\partial \hat r })^{f_m} ,
\end{equation}
where $a_i , b_i , c_j , d_j \in {\bf N }_o , \  i\in \{ 1, \cdots , n
\}, j\in \{1,\cdots , m \}, \  n\in {\bf N} ,\ m=\sum _i a_i + \sum _j
b_j +1 , \  e_i ,e_j \in \{ 0,1\}$ and $\hat r \in \{ \hat q , \hat p
\}$. Modifications of the other properties follow straightforwardly. On
the other hand, it is not needed to modify above expressions for the
identity operator because it can be seen as $\hat q ^o$ or $\hat p ^o$.

The expression (4) should be taken, we propose, as corresponding to the
monomial $q^n p^m$ of classical mechanics and, since $\hat q ^n \circ
\hat p ^m$ should represent the symmetrized product of $n$ operators of
coordinate and $m$ operators of momentum, it was constructed from these
$n+m$ operators in two steps (the first was to treat these operators
as ordinary numbers, the meaning of which is to take $\hat q ^n \hat p
^m$, or any other sequence of these $n+m$ operators, and the second
step was to act with ${\bf S}$ on that). Then, by applying this
procedure to a bit complicated case, one can easily answer the question
what shall be the symmetrized product of two monomials, let say, $\hat
q ^a \circ \hat p ^b$ and $\hat q ^c \circ \hat p ^d$ (both monomials
are symmetrized in the above way). Namely, the result should be the
expression gained after the expressions similar to (4) were multiplied
as it would be done in the c-number case and after ${\bf S}$ was
applied on that. After the first step, one would get:
$$
{a!b!\over (a+b)!}{c!d!\over (c+d)!} (\hat q ^a \hat p ^b \hat q ^c
\hat p ^d + \cdots + \hat p ^b \hat q ^a \hat p ^d \hat q ^c ),
$$
where, in the parenthesis, there should be ${(a+b)!\over
a!b!}{(c+d)!\over c!d!}$ terms on which ${\bf S}$ acts in the same
way. The final result would be:
$$
{(a+c)!(b+d)! \over (a+b+c+d)!}(\hat q ^{a+c} \hat p ^{b+d} + \cdots +
\hat p ^{b+d} \hat q ^{a+c} ).
$$
Again, in the parenthesis, there should stand only different
combinations of $a+b+c+d$ operators, where $a+c$ are of the one kind
and $b+d$ are of the other. In more compact notation, the above reads:
$$
(\hat q ^a \circ \hat p ^b ) \circ (\hat q ^c \circ \hat p ^d )={\bf
S}({\bf S}(\hat q ^a \cdot \hat p ^b )\cdot {\bf S}(\hat q ^c \cdot
 \hat p ^d ))= \hat q ^{a+c} \circ \hat p ^{b+d} .
$$
Consequently, the symmetrized product of quantum mechanical observables
$\sum _i c_i \hat q ^{n_i} \circ \hat p ^{m_i}$ and $\sum _j d_j \hat q
^{r_j} \circ \hat p ^{s_j}$ should be:
\begin{equation}
(\sum _i c_i \hat q ^{n_i} \circ \hat p ^{m_i}) \circ (\sum _j d_j \hat
 q ^{r_j} \circ \hat p ^{s_j}) = \sum _i \sum _j c_i d_j  \hat q ^{n_i
+ r_j} \circ \hat p ^{m_i + s_j}.
\end{equation}
The involved monomials are defined above.

\section{Some properties of the symmetrized product}

Our proposal of the symmetrized product is of the same complexity
at the level of correspondence principle as are those given in
literature: ordinary product, multiplication by numbers and
summation all have to be used in order to define it.

Our proposal differs in that we find it necessary to look on the
symmetrized product as on a two step operation. If it is so when
the correspondence principle is under consideration, then we find
it necessary to apply both of the mentioned steps in all other
situations where the algebraic structure of quantum mechanics is
addressed.

If one multiplies in ordinary manner operatorial expressions that
were symmetrized in either way, then it does not come as surprise
that result is no longer adequately ordered. Moreover, the
omission at this place of the second step - the application of
${\bf S}$, makes such approaches inconsistent. That is because
basic operatorial polynomials, which by definition correspond to
classical mechanical variables, actually are symmetrical products
of several $\hat q$'s and
$\hat p$'s introduced at the beginning of quantization.

However, the ordered product is not always applied in the standard
quantum mechanics, see Messiah (1961). For example, the
calculation of dispersion asks to take ordinary product of
operators instead of any of the proposed algebraical ones. This
implicates that one has to be more specific in writing, let say,
$f(\hat q , \hat p )^2$ since it is not obvious is it $f(\hat q ,
\hat p ) \cdot f(\hat q , \hat p )$ or $f(\hat q , \hat p )\circ
f(\hat q , \hat p )$. There are situations where one needs the
successive application of operators on states being uninterested
in the algebraic aspect of quantum mechanics and {\it vice versa}.
Needless to say, the last two expressions differ in general.

The algebra of quantum mechanical observables will be formally
closed and there will be no purely algebraic obstructions with
$\circ$ because the symmetrized product is a two step operation in
all instances (when the correspondence with classical variables is
established and when those quantum observables are multiplied) and
because ${\bf S}$ is defined for sequences of $\hat q$'s and $\hat p$'s
on the first place and not for operators in general like it is the case
in Temple (!935). This means that, for the calculation of the product
of general observables, it is necessary to express them like it was
done on the LHS of (10). These, we believe, guarantee that one will not
go out of the proposed way of symmetrization. Since all combinations of
involved $\hat q$ and $\hat p$ appear on the RHS of (10), the
symmetrized product of two symmetrized polynomials, which are the
Hermitian, obviously will be invariant under the Hermitian conjugation.

It is necessary to declare the action of $\bf S$ on $\hbar$ for it
is dimensional constant, not a complex number. Metaphorically
speaking, in contrast to the other constants it ``incorporates''
coordinate and momentum whose numbers of appearances in
expressions are important from the point of view of $\bf S$. The
property (8) prevents one to end with some ambiguity in case when
the commutation relation $[\hat q  , \hat p ]=i\hbar \hat I$ is
used for the reexpression of operatorial sequences before the
application of $\bf S$. For example, without (8), the symmetrized
product of ${1\over 2}(\hat q \hat p + \hat p \hat q )$ with
itself will not be equal to the symmetrized product of $\hat q
\hat p - {i\hbar \over 2}$ with itself, while ${1\over 2}(\hat q
\hat p + \hat p \hat q )=\hat q \hat p - {i\hbar \over 2}$. In
abstract considerations we are proceeding, the Planck constant can
appear only due to the above commutation relation; we ignore
artificial situations when $\hbar$ is introduced in expressions by
hand. It could be said that, being the commutant of coordinate and
momentum, $\hbar$ is the quantum analogue of zero - the commutant
of classical  $q$ and $p$. And, since zero annihilates all
sequences of $q$ and $p$ in $\cdot$ product, the Planck constant
should do the same for sequences of $\hat q$ and $\hat p$ in
$\circ$ product.

Another similarity between $\circ$ and $\cdot$ is that they are
commutative. This characteristic of proposed symmetrized product
follow directly from the fact that for the symmetrizer
$\bf S$ only the numbers of $\hat q$'s and $\hat p$'s in a
sequence on which it acts are relevant, in contrast to that how
they are ordered in the sequence. These make us to believe that
$\circ$ is defined in such a way to be the complete imitation of the
product of classical mechanics.

Next remark is related to the situations where the form of
operatorial expressions is fixed for some reasons. In Messiah (1961) it
was said that ordered product of $f(\hat {\vec r})$ and $\hat {\vec p}$
has to be the one half of the anti-commutator of these two. Adapted to
the present considerations, we have to show that:
\begin{equation}
{1\over 2}(\sum _{n=o} ^N c_n \hat q ^n  \hat p + \hat p \sum _{n=o} ^N
c_n \hat q ^n )=\sum _{n=o} ^N c_n \hat q ^n \circ \hat p ,
\end{equation}
since, with our proposal of symmetrized product, we do not want to
contradict the well-known facts of standard quantum mechanics. In order
to do that, we firstly transform the LHS of (11) into $\sum _{n=o} ^N
c_n \hat q ^n  \hat p  - {i\hbar \over 2} {\partial \over \partial \hat
q}\sum _{n=o} ^N c_n \hat q ^n $ and then reexpress each term on
the RHS of (11) in the following way:
$$
\sum _{n=o} ^N c_n {1\over {n+1}}(\hat p \hat q ^n + \hat q \hat p \hat
q ^{n-1} +\cdots + \hat q ^{n-1} \hat p \hat q + \hat q ^n  \hat p )=
$$
$$
=\sum _{n=o} ^N c_n {1\over n+1}(\hat q ^n \hat p -i\hbar n \hat q
^{n-1} + \hat q (\hat q ^{n-1} \hat p -i\hbar (n-1) \hat q ^{n-2})
+\cdots +\hat q ^{n-1}(\hat q \hat p -i\hbar )+
$$
$$
+\hat q ^n \hat p )=\sum _{n=o} ^N c_n \hat q ^n \hat p -i\hbar \sum
_{n=o} ^N c_n {1\over {n+1}}(n + (n-1) + \cdots +1) \hat q ^{n-1}=
$$
$$
=\sum _{n=o} ^N c_n \hat q ^n \hat p - {i\hbar \over 2} \sum _{n=o} ^N
c_n n \hat q ^{n-1} =
$$
$$
=\sum _{n=o} ^N c_n \hat q ^n \hat p - {i\hbar \over 2} {\partial
\over \partial \hat q} \sum _{n=o} ^N c_n \hat q ^n .
$$
So, our proposal of symmetrized product is not in conflict with
the mentioned demand since both sides of (11) are equal. Up to our
knowledge, this is the only situation where the form of observable is
{\it a priori} fixed.

Finally, one can show that:
\begin{equation}
{\partial \over \partial \hat q} (\hat q ^n \circ \hat p ^m )= n \hat q
^{n-1} \circ \hat p ^m,
\end{equation}
the meaning of which is that the proposed algebraic structure of
quantum mechanics is formally closed under the partial derivations
- the type of symmetrization is saved under the action of
${\partial \over
\partial \hat q}$ and ${\partial \over \partial \hat p}$. This will be
important for redefinition of the Lie bracket of quantum
mechanics that we are going to discuss in the next section.

For this purpose, one has to show that, after the partial derivation of
$\hat q ^n \circ \hat p ^m$, there will be the sum of all ${(n-1+m)!\over
(n-1)!m!}$ different combinations of $n-1$ operators of coordinate and
$m$ operators of momentum and that each of these combinations will
appear $n+m$ times, so the multiplicative factor ${n!m!\over
(n+m)!}$, standing in front of the sum and  coming from $\hat q ^n
\circ \hat p ^m$, will be regularized.

It is trivial fact that, after the partial derivation of sequences of
$n$ operators of coordinate and $m$ operators of momentum (that appear
in $\hat q ^n \circ \hat p ^m$), there will be sequences of $n-1$
operators of coordinate and $m$ operators of momentum (which is asked
by $\hat q ^{n-1} \circ \hat p ^m$). After the partial derivation of
$\hat q ^n \circ \hat p ^m$, there will be the sum of all different
combinations of $n-1$ operators of coordinate and $m$ operators of
momentum since, by definition, in $\hat q ^n \circ \hat p ^m$, all
different combinations of appropriate operators are present. Each
of these ${(n+m-1)!\over (n-1)!m!}$ different combinations needed
for $\hat q ^{n-1} \circ \hat p ^m$ will follow after the derivation of
many different combinations - sequences, appearing in $\hat q ^n \circ
\hat p ^m$. All those sequences of $\hat q ^n \circ \hat p ^m$, that
will give after derivation a particular sequence of $\hat q ^{n-1}
\circ \hat p ^m$, have one more $\hat q$. This ``extra'' $\hat q$,
that is going to be annihilated by derivation, can be on $n+m$
different places when those sequences are compared to the chosen
sequence of $\hat q ^{n-1} \circ \hat p ^m$. (In other words, the
sequences of $\hat q ^n \circ \hat p ^m$ that will be transformed
by ${\partial \over \partial \hat q}$ in the  chosen one differ
from it only in one $\hat q$, all other $\hat q$'s and $\hat p$'s are
equally ordered. To visualize this, it is helpful to consider, for
instance, $\hat q ^{n-1} \hat p ^m$; it follows after derivation of:
$\hat q ^n \hat p ^m$, $\hat q ^{n-1} \hat p \hat q \hat p ^{m-1}$,
$\cdots$, $\hat q ^{n-1} \hat p ^{m-1} \hat q \hat p$  and $\hat q
^{n-1} \hat p ^m \hat q$ that are present in $\hat q ^n \circ \hat p
^m$. The mentioned extra $\hat q$ in $\hat q ^n \hat p ^m$ can be each
of $n$ operators of coordinate, like it appears on $n$  different
places. In the next $m-1$ sequences, it is the $\hat q$ among
operators of momentum that is extra and, in the last combination, it
$\hat q$ standing on the right of $\hat p ^m$.) This means that each
sequence of $\hat q$'s and $\hat p$'s needed for $\hat q ^{n-1} \circ
\hat p ^m$ will follow $n+m$ times after the application of ${\partial
\over \partial \hat q}$ on $\hat q ^n \circ \hat p ^m$. In this
way one can convince oneself that (12) holds and, by proceeding in
similar manner for the partial derivation with respect to the
momentum and in the case of polynomials, one can show that the
algebra of observables is formally closed under the action of
${\partial \over \partial \hat q}$ and ${\partial \over \partial
\hat p}$.

\section{ The redefined Lie bracket of quantum mechanics }

The symmetrized product we have proposed can be used in the
classical mechanics instead of the standard algebraic
multiplication of variables since the ordering rules have no
effect in the case of the commutative algebra. So, one can say
that the algebraic product of both mechanics is one and the same.
Then, the difference between them is in the Lie bracket and, in
what follows, we shall try to remedy this by intervening on this
bracket of quantum mechanics. We propose the substitution of the
commutator divided by $i\hbar$ with the operatorial form of
Poisson bracket:
\begin{equation}
\{ f(\hat q , \hat p ), g(\hat q , \hat p  )\} _{\bf S} = {\partial
f(\hat q , \hat p )\over \partial \hat q } \circ  {\partial g(\hat q , \hat
p )\over \partial \hat p } -  {\partial g(\hat q , \hat p )\over
\partial \hat q }\circ {\partial f(\hat q , \hat p )\over \partial \hat p
}.
\end{equation}
{\it Nota bene}, in difference to (1), the operators are involved in the
last expression and symmetrized product stays instead of the ordinary
one.

Before one addresses the obstruction to quantization mentioned in
Sec. 1 and before reexpresses dynamical equation of quantum
mechanics, one has to convince oneself that the just defined
``symmetrized'' Poisson bracket has all properties of the Lie
product. To do this, one can proceed as follows. That $\{ \ \ ,\ \
\} _{\bf S}$ is linear one easily concludes remembering that the
partial derivations and symmetrized product are linear operations.
That it is anti-symmetric one can see directly from definition. By
comparing the operatorial expression under consideration with the
adequate one of classical mechanics, one finds that the partial
derivations and symmetrized products of symmetrized operatorial
expressions within $\{ \ \ ,\ \ \} _{\bf S}$ in complete imitate
the partial derivations and ordinary products of c-number
functions in $\{ \ \ ,\ \ \} $. So, the confirmation of the Jacoby
identity rests on the analogy - each step of the calculation in
the case of operators has the corresponding one in the c-number
case.

Due to mentioned analogies, the symmetrized Poisson bracket would behave as
derivative:
$$
\{ f(\hat q , \hat p ) , g(\hat q , \hat p ) \circ h(\hat q , \hat p ) \}
_{\bf S} =
$$
\begin{equation}
=\{ f(\hat q , \hat p ) , g(\hat q , \hat p ) \} _{\bf S} \circ h(\hat
q , \hat p ) + g(\hat q , \hat p ) \circ \{ f(\hat q , \hat p ), h(\hat
q , \hat p ) \} _{\bf S},
\end{equation}
This can be checked by taking three monomials in $\hat q$ and $\hat p$.
For $f(\hat q , \hat p) = \hat q ^{a_1} \circ \hat p ^{a_2}$, $g(\hat q
, \hat p) = \hat q ^{b_1} \circ \hat p ^{b_2}$ and $h(\hat q , \hat p)
= \hat q ^{c_1} \circ \hat p ^{c_2}$, both sides of (14) will be $(a_1
(b_2 +c_2 ) - a_2 (b_1 +c_1 ))\hat q ^{a_1 + b_1 +c_1 -1}  \circ\hat p
^{a_2 +b_2 + c_2 - 1}$.

If one takes ordinary multiplied $g(\hat q ,\hat p )$ and $h(\hat
q ,\hat p )$, then: $$ \{ f(\hat q , \hat p ) , g(\hat q , \hat p
) \cdot h(\hat q , \hat p ) \} _{\bf S} \ne $$ $$ \{ f(\hat q ,
\hat p ) , g(\hat q , \hat p ) \} _{\bf S} \cdot h(\hat q , \hat p
) + g(\hat q , \hat p ) \cdot \{ f(\hat q , \hat p ), h(\hat q ,
\hat p ) \} _{\bf S}, $$ in general. One and the same product has
to be used for all multiplications in $\{ \ \ ,\ \ \} _{\bf S}$
for the Leibniz rule to be satisfied.

The generalization of the symmetrized Poisson bracket for more than one
degree of freedom is influenced by requirements coming from physics,
see Prvanovi\'c and Mari\'c (2000) and references therein. This topic
we shall consider in the forthcoming article belonging to the series
concerned with the foundation of theory of hybrid systems.

Given remarks make trivial the problem of possibility for \
obstruction to quantization based on the symmetrized \ Poisson
bracket as the Lie product. If there was some equation for
classical variables, then the same equation will hold for their
quantum counterparts since the symmetrized product and the
symmetrized Poisson bracket imitate the adequate operations in
c-number case. Said in more descriptive way, the equation in
quantum mechanics will hold since it differs from the
corresponding equation of classical mechanics only in that there
are hats above coordinate and momentum and there is $\circ$
instead of
$\cdot$. Consequently, this quantization is, we believe, unambiguous,
{\it i.e.}, obstruction free {\it in toto}.

Can the symmetrized Poisson bracket substitute the commutator in von
Neumann equation (3) is the last question that we are going to address.
Related to this are: how the operators representing states of quantum
system can be expressed as depending on some functions of $\hat q$ and
$\hat p$ and how should the symmetrizer ${\bf S}$ act on sequences
formed of $\hat q$, $\hat p$ and partial derivatives of the state $\hat
\rho$. With the help of:
\begin{equation}
\vert q' \rangle \langle q' \vert = \int \delta (q-q')\vert q
\rangle \langle q \vert dq = \delta (\hat q - q'),
\end{equation}
and:
\begin{equation}
\vert q'' \rangle \langle q' \vert = e^{{1\over i\hbar} (q'' - q')\hat
p} \vert q' \rangle \langle q' \vert = e^{{1\over i\hbar} (q'' -
q')\hat p } \cdot  \delta (\hat q - q'),
\end{equation}
one immediately finds that the pure state in general, {\it i.e.},
$\vert \psi \rangle = \int \psi (q) \vert q \rangle dq$, can be
expressed as:
$$
\vert \psi \rangle \langle \psi \vert = \int \int \psi (q) \psi ^* (q')
\vert q \rangle \langle q' \vert dq dq' =
$$
\begin{equation}
=\int \int \psi (q) \psi ^* (q') e^{{1\over i\hbar} (q'' - q')\hat p}
\cdot \delta (\hat q - q') dq dq'.
\end{equation}
But, $\delta (\hat q - q')$ is neither polynomial nor analytical
function of $\hat q$. This means that it can not be expressed in
the form $\sum _n c_n \hat q ^n$ which is necessary according to
(5) for a direct calculation of the symmetrized product.
Supplementary defining property of the ordering rule for a novel
situation, when $\bf S$ acts on sequences of $\hat q$, $\hat p$
and the partial derivatives of $\delta (\hat q - q')$ and/or the
general state, are certainly needed. This property is (9) the meaning
of which is that the partial derivatives of states are entities
different from $\hat q$ and $\hat p$.

If the symmetrical product of $\hat q ^n \circ \hat p ^m$ and partial
derivative of $\vert \psi \rangle \langle \psi \vert$ is to be
calculated, one should look for ${(n+m+1)!\over n! m!}$ distinct
combinations of involved operators. Concretely:
\begin{equation}
\hat p  \circ {\partial \vert \psi \rangle \langle \psi \vert \over
\partial \hat q } = {1\over 2} (\hat p {\partial \vert \psi \rangle
\langle \psi \vert \over \partial \hat q } + {\partial \vert \psi
\rangle \langle \psi \vert \over \partial \hat q } \hat p ),
\end{equation}
and:
$$
\hat q ^n \circ {\partial \vert \psi \rangle \langle \psi \vert \over
\partial \hat p } =
$$
\begin{equation}
{1\over n+1} (\hat q ^n {\partial \vert \psi \rangle
\langle \psi \vert \over \partial \hat p } + \hat q ^{n-1} {\partial
\vert \psi \rangle \langle \psi \vert \over \partial \hat p } \hat q +
\cdots + \hat q {\partial \vert \psi \rangle \langle \psi \vert \over
\partial \hat p } \hat q ^{n-1} + {\partial \vert \psi \rangle \langle
\psi \vert \over \partial \hat p } \hat q ^n ).
\end{equation}

Then, one can easily show that:
\begin{equation}
\{ {\hat p ^2 \over 2m} + V(\hat q) , \vert \psi \rangle \langle \psi
\vert \} _{\bf S} = {1\over i\hbar } [ {\hat p ^2 \over 2m} + V(\hat q
) , \vert \psi \rangle \langle \psi \vert ],
\end{equation}
where $V(\hat q )=\sum _n c_n \hat q ^n$. Instead of proving (20) in
the coordinate representation, one can simply substitute ${\partial
\over \partial \hat q }$ and ${\partial \over \partial \hat p }$ that
appear on the LHS of (20) with ${1\over i\hbar } [\  , \hat p ]$ and
${1\over i\hbar }[\hat q ,\ ]$, respectively, and, by using (18) and
(19), after few elementary steps one will find that the LHS and RHS of
(20) are equal. Despite of being less interesting for physics, let us
pay more attention on:
\begin{equation}
\{ F(\hat q , \hat p ) , \vert \psi \rangle \langle \psi \vert \} _{\bf
S} = {1\over i\hbar } [ F(\hat q , \hat p ) , \vert \psi \rangle
\langle \psi \vert ],
\end{equation}
where $F(\hat q , \hat p )$ is the general element of the quantum
mechanical algebra. The analysis of (21) can start with
considerations of some monomial $\hat q ^n \circ \hat p ^m$. In
order to obtain the LHS of (21), according to (12), one,
firstly, has to multiply symmetrically $n \hat q ^{n-1} \circ \hat
p ^m$ and ${\partial \hat \rho \over \partial \hat p}$, where
$\hat \rho = \vert \psi \rangle \langle \psi \vert $, then to
multiply symmetrically $m \hat q ^n \circ \hat p ^{m-1}$ and
$-{\partial \hat \rho \over \partial \hat q}$ and, finally, to add
these two. Due to (9), the LHS of (21) then becomes:
\begin{equation}
{n! m! \over (n+m)!} (\hat q ^{n-1} \hat p ^m {\partial \hat \rho
\over \partial \hat p} + \cdots + {\partial \hat \rho \over
\partial \hat p} \hat p ^m \hat q ^{n-1} - (\hat q ^n \hat p ^{m-1}
{\partial \hat \rho \over \partial \hat q} + \cdots + {\partial
\hat \rho \over \partial \hat q} \hat p ^{m-1} \hat q ^n )).
\end{equation}
After substituting ${\partial \hat \rho \over \partial \hat p}$ with
${1\over i\hbar} [ \hat q , \hat \rho ]$ and ${-\partial \hat \rho \over
\partial \hat q}$ with ${1\over i\hbar} [ \hat p , \hat \rho ]$, one
should look how to simplify (22). Some terms in (22) will be of the form
$\hat A \hat q (\hat q \hat \rho ) \hat q \hat B$, where $\hat A$ and
$\hat B$ represent (different) sequences of $\hat q$'s and $\hat p$'s
and $(\hat q \hat \rho )$ means that these two come from the
commutator $[ \hat q , \hat \rho ]$. But, such terms will be
canceled by $-\hat A \hat q \hat q (\hat \rho  \hat q ) \hat B$
(which certainly should appear in (22)), where the minus sign
comes from the commutator. Proceeding in this way for all other
forms, one can find that many terms in (22) will mutually cancel
each other. This holds for all terms except those where $\hat
\rho$ stands at the beginning or at the end of the sequence.
Consequently, (22) is equal to:
$$
{1\over i\hbar } {n! m! \over (n+m)!} ((\hat q ^n \hat p ^m + \cdots +
\hat p ^m \hat q ^n ) \hat \rho  - \hat \rho (\hat q ^n \hat p ^m +
\cdots + \hat p ^m \hat q ^n )),
$$
which is nothing else than the RHS of (21) for the considered monomial.
Then, due to the linearity of commutator and symmetrized Poisson
bracket, (21) will hold for general $F(\hat q , \hat p )$, too.

It should be noticed that (20) will not hold for $V(\hat q )=\hat q
^n$ when $n\ge 3$ if the symmetrized product within Poisson bracket is
taken to be the one half of the anti-commutator of involved operators.
This is why we have defined $\bf S$ to be the superoperator that
produces (19), {\it i.e.}, defined by (9).

From (20) and (21), it follows that the dynamical equation of
quantum mechanics can be reexpressed. Obviously, resulting
equation is operatorial version of the Liouville equation:
\begin{equation}
{\partial \rho (\hat q ,\hat p , t) \over \partial t }= \{ H(\hat q ,
\hat p ), \rho (\hat q ,\hat p ,t) \} _{\bf S}.
\end{equation}
It is understood that the Hamiltonian is $H(\hat q , \hat p )=\sum _i
c_i \hat q ^{n_i} \circ \hat p ^{m_i}$.

\section{Concluding remarks}

For the introduction of the symmetrized product that we have
proposed it was necessary to look on the operators of coordinate
and momentum as on the basic elements. The special position that
coordinate and momentum have in mechanics has manifested itself in
that all physically meaningful entities have to be defined, or
expressed, via them. The algebraic and Lie algebraic products have
to be defined, on the first place, with respect to the coordinate
and momentum and not for the operators in general the particular
examples of which would be $\hat q$ and $\hat p$. Only after
observables are expressed in form of the functions of $\hat q$ and
$\hat p$, then their algebraic or Lie algebraic product can be
found. The numbers of appearances of $\hat q$'s and $\hat p$'s in
operatorial sequences play the crucial role in such situations,
just like degrees of $q$ and $p$ do in the c-number case.

As we have stressed, the only way to make the approach consistent
and, we believe, to avoid ambiguities is to apply the same
ordering procedure in all situations. If it was seen as a two step
operation at the level of the correspondence principle, then it
should be treated in the same manner in other occasions as well.
And, moreover, we believe that the ordering procedure has to be
taken as a two step operation if the Hermiticity of observables is
requested.

The unavoidable second step of all multiplications, {\it i.e.},
the symmetrizer $\bf S$, due to properties that it produces all
distinct combinations of coordinate and momentum and annihilates
$\hbar$, makes the approach obstruction free. When the operators
introduced at the beginning of quantization are looked from the
point of view of either algebra or Lie algebra, there will be no
contradictory statements because $\bf S$ within $\circ$ and $\{ \
\ , \ \ \} _{\bf S}$ forms these two operations to be the complete
imitation of the corresponding ones of classical mechanics.

At last, let us summarize in brief our proposition for the
quantization. The basic elements of classical mechanical algebra: $q$,
$p$ and $1$, should be mapped into: $\hat q$, $\hat p$ and $\hat I$,
respectively, where $[\hat q , \hat p ] = {1\over i\hbar} \hat I$.
The general element $\sum _i c_i q^{n_i} \cdot p^{m_i}$ should be
mapped in $\sum _i c_i \hat q ^{n_i} \circ \hat p ^{m_i}$. The
ordinary product of $f(q,p)=\sum _i c_i q^{n_i} \cdot p^{m_i}$ and
$g(q,p)=\sum _j d_j  q^{r_j} \cdot p^{s_j}$ should be translated
in (10), and their Lie product (1) should become (13). The dynamical
equations for states, which are expressible via coordinate and momentum
in both mechanics, should be (2) in classical mechanics and (23) in
quantum mechanics.

If we neglect for the moment that classical mechanics is
represented in this article with c-number functions, while quantum
mechanics uses operatorial functions, then it could be said not
only that classical and quantum mechanics are equivalent regarding
the mathematical structures, but the algebraic and Lie algebraic
products have the same realization for both mechanics since the
symmetrized product and symmetrized Poisson bracket can be used
without any problem instead of the ordinary ones in classical
mechanics, as well. This we are going to use in the next article
devoted to the formulation of theory of hybrid systems.

\end{document}